\documentclass[aps,prl,twocolumn,floats,preprintnumbers]{revtex4}
\usepackage{graphicx}
\usepackage{hyperref}
\usepackage{times}
\begin{document}

\preprint{FERMILAB-Pub-03/002-T}
\preprint{UCB-PTH-02/62}
\preprint{LBNL-51969}

\title{Statistical Test of Anarchy}

\author{Andr\'e de Gouv\^ea$^{1}$ and Hitoshi Murayama$^{2,3}$}
\affiliation{$^1$ Theoretical Physics Division, Fermilab, P.O. Box
  500, Batavia, IL~60510-0500, USA} 

\affiliation{$^2$Department of Physics, University of California,
  Berkeley, CA~94720, USA}

\affiliation{$^3$ Theoretical Physics Group, Lawrence Berkeley
  National Laboratory, Berkeley, CA~94720, USA} 

\date{\today}

\begin{abstract}
  ``Anarchy'' is the hypothesis that there is no fundamental
  distinction among the three flavors of neutrinos.  It describes the
  mixing angles as random variables, drawn from well defined
  probability distributions dictated by the group Haar measure.  We
  perform a Kolmogorov--Smirnov (KS) statistical test to verify
  whether anarchy is consistent with all neutrino data, including the
  new result presented by KamLAND.  We find a KS probability for
  Nature's choice of mixing angles equal to 64\%, quite consistent with
  the anarchical hypothesis.
  In turn, assuming that anarchy is indeed correct, we compute
  lower bounds on $|U_{e3}|^2$, the remaining unknown ``angle'' 
  of the leptonic mixing matrix.  
  \end{abstract}

\maketitle

All fermions in the Standard Model of particle physics (SM) seem to
come in threes. The three copies of each fundamental matter particle
have in common all properties except one -- the mass.  It is common to
say that there are three families, generations, or {\it flavors}\/ of
each matter particle in the SM.  Currently we do not know the reason
behind the number three, nor why the matter particles should
``repeat'' at all.  Therefore, it is important to look for any
information that may shed light into the origin of flavor.

Within the SM, it has been known for quite some time that different
quark flavors can mix quantum mechanically, and that the weak
interactions can turn one flavor into another. The ``amount'' of
mixing is summarized by the so-called Cabibbo--Kobayashi--Maskawa
(CKM) unitary matrix. The CKM matrix, in turn, can be parameterized by
three mixing angles $\theta_{12}, \theta_{13}, \theta_{23}$ and one
complex phase $\delta$ (throughout, we use the ``PDG parameterization''
\cite{PDG} for the mixing matrices).  A non-vanishing phase $\delta$
indicates that SM processes can violate CP invariance, distinguishing
matter from anti-matter in a subtle manner. With the beautiful data
from the $B$-factory experiments, we have been able to confirm the CKM
framework, and measure all angles and the CP-odd phase with $O(10)\%$
accuracy.  

A noteworthy feature of the CKM matrix is that it is rather well
approximated by the unit matrix, meaning that the quark mixing angles
are all small. This fact, combined with the fact that the quark masses
are quite distinct (the ratio of the lightest to
heaviest quark mass is $O(10^{-5})$), is interpreted as evidence for
the existence of some underlying symmetry or physical mechanism that
differentiates the quark families and hence explains the hierarchy
in the quark masses and the small mixing angles.

In the SM, all neutrinos are exactly massless. This being the case,
one can always choose a basis where the Maki--Nakagawa--Sakata (MNS)
unitary matrix, the leptonic analog of the CKM matrix, is the unit
matrix without loss of generality. This means that there are no SM
processes through which one lepton flavor can turn into another.  This
hypothesis has been indeed confirmed by all experimental searches for
charged lepton flavor violation to date \cite{PDG}.

If neutrinos have masses, and these masses are distinct, there is no
reason to expect that the MNS matrix is trivial, and lepton flavor
transitions are observable in principle. In this case, 
the most sensitive probes for
lepton flavor transitions are neutrino oscillation processes, through
which a neutrino produced in a well-defined flavor state
$\nu_{\alpha}$ is detected in a different flavor state $\nu_{\beta}$
after propagating over a macroscopic distance $L$. The transition
probabilities depend on the mixing angles and the CP-odd phase of the
MNS matrix, plus the difference of the neutrino masses-squared, $\Delta
m^2_{ij}\equiv m_i^2-m_j^2$.

Since 1998, there is compelling evidence that neutrino flavor
transitions do occur when the neutrinos traverse macroscopic
distances. Atmospheric \cite{atm}, solar \cite{solar}, and, very
recently, reactor neutrino experiments \cite{KamLAND} have all
observed data consistent with the neutrino oscillation hypothesis. In
light of all the experimental evidence, it appears that neutrinos have
masses, and that leptonic flavors mix.

There are two striking features regarding the values of the
oscillation parameters which are extracted from the current
neutrino data. One is that the neutrino masses are extremely small.
Neutrino oscillation experiments have determined that the neutrino
mass-squared differences are \footnote{We define
  the neutrino mass eigenvalues such that $m_2^2>m_1^2$, and $\Delta
  m^2_{12}<|\Delta m^2_{13,23}|$.} $|\Delta
m^2_{23}|=(2-7)\times10^{-3}$~eV$^2$ \cite{atm} and $\Delta
m^2_{12}=(4-20)\times 10^{-5}$~eV$^2$ \cite{KamLAND}. 
These results, combined with
direct searches for neutrino masses \cite{PDG}, yield that the
heaviest neutrino mass is less than $O(1)$~eV, over six orders of
magnitude smaller than the smallest charged fermion mass of which we
know (the electron mass). The other is that, of
the mixing angles, two ($\theta_{12},\theta_{23}$) are known with some
precision, and are both large: $\sin^22\theta_{23}\gtrsim0.9$
\cite{atm} and $\sin^22\theta_{12}\gtrsim0.4$ \cite{KamLAND}.

Assuming a three family mixing scenario, there are two more parameters
in the MNS mixing matrix that are still unknown: $\theta_{13}$ and
$\delta$. In particular, if $\delta\neq 0$ neutrino oscillation
processes need not conserve CP.  Leptogenesis models
\cite{leptogenesis}, on the other hand, 
try to relate the existence of matter but no
anti-matter in the Universe to the CP violation present in the
neutrino sector, making its observation of the utmost 
interest.  CP-violating effects parameterized by the CP-odd phase 
$\delta$ of the MNS matrix can be probed in accelerator-based long-baseline 
neutrino oscillation experiments if, for example, one compares the flavor 
transformation probabilities of neutrinos and anti-neutrinos 
(written here in vacuum),
\begin{eqnarray}
  & & P(\nu_e \rightarrow \nu_\mu) - P(\bar{\nu}_e \rightarrow
  \bar{\nu}_\mu) = -16 s_{12} c_{12} s_{13} c^2_{13} s_{23} c_{23}
  \nonumber \\
  & & \qquad \sin \delta \sin \frac{\Delta m^2_{12}L}{4E} 
  \sin \frac{\Delta m^2_{13}L}{4E} \sin \frac{\Delta m^2_{23}L}{4E}\ ,
\end{eqnarray}
where $s_{ij} = \sin \theta_{ij}$, $c_{ij} = \cos \theta_{ij}$.  It is
well known that the observation of CP violation in neutrino
oscillations is possible only if $\theta_{12}$ and $\Delta m^2_{12}$
are ``large enough'' (and the atmospheric parameters are also large, as
has been established by the atmospheric data). The KamLAND result has
shown that this is the case.  The remaining question, therefore, is
whether $\theta_{13}$ is also large enough to render the experimental
search for CP violation possible.  The only information we currently
have is that $\theta_{13}$ is relatively small:
$\sin^2\theta_{13}\lesssim 0.05$, constrained by the CHOOZ experiment 
\cite{CHOOZ}.

The purpose of this letter is two-fold.  First, we examine if the
current data ``requires'' new symmetry principles in order to control
the structure of the MNS matrix, analogous to the situation in the
quark sector.  Saying that there is no symmetry principle behind the
MNS matrix means there is no fundamental distinction among the three
flavors of neutrinos.  If this is the case, the MNS matrix is
distributed (statistically) according to the bi-invariant Haar measure
of group theory, which dictates the probability distribution of the
mixing angles.  The hypothesis here is that Nature has chosen one
point according to this probability distribution.  This is the concept
of ``anarchy'' in neutrinos \cite{anarchy1,anarchy2}.  
%Given the
%impressive progress of our understanding of the neutrino sector, 
We would like to examine if the data are consistent with anarchy by
performing a Kolmogorov--Smirnov (KS) statistical test.  We find that they
are perfectly consistent. 
%\footnote{Some authors have claimed that the
%neutrino sector prefers to have a symmetry principle behind their
%mass and mixing patterns as well, see, {\it e.g.}\/, 
%V.~Antonelli, F.~Caravaglios, R.~Ferrari and M.~Picariello,
%%%``Neutrino masses and non Abelian horizontal symmetries,''
%Phys.\ Lett.\ B {\bf 549}, 325 (2002);
%%%CITATION = HEP-PH 0207347;%%
%G.~Altarelli, F.~Feruglio and I.~Masina,
%%%``Models of neutrino masses: Anarchy versus hierarchy,'' hep-ph/0210342.
%%%CITATION = HEP-PH 0210342;%%
%These authors appear to disregard the fact that the ratio $\Delta
%m^2_{12}/\Delta m^2_{23}$ is predicted successfully in the anarchy
%with the seesaw mechanism.}.

Second, given the empirical success of anarchy, we study what it has
to say about $\theta_{13}$.  Anarchy prefers large values for
$\theta_{13}$, meaning that a small $\theta_{13}$ would be
inconsistent with the anarchical hypothesis.  By turning this argument
around, we can place a {\it lower limit on}\/ $\theta_{13}$ at various
confidence levels, again using the KS test.

%{\sl The Kolmogorov Smirnov Statistical Test}--
Consider the following situation: there is a model that ``predicts''
that a certain quantity is described by a probability
distribution. For example, one may construct a model that predicts
that a given quantity $x$ may have any value from 0 to 1, with equal
probability. This means that the probability density $f(x)$ is
\footnote{The probability density function is defined in such a way
  the probability that $x$ has a value between $x_0$ and $x_0+{\rm
    d}x$ is given by $f(x_0){\rm d}x$.}
\begin{equation}
  \label{f(x)}
  f(x) = \left\{ 
    \begin{array}{ll} 
      1 & {\rm if}~x\in [0,1], \\
      0 & {\rm otherwise.} 
    \end{array} \right. 
\end{equation}

Let us assume that the value of $x$ is known: $x=x_0$.  The question
to be addressed is how well does the result $x=x_0$ agree with the
model presented above (that the probability density for $x$ is given
by Eq.~(\ref{f(x)}))?  This question can be answered using the KS
test.  Given that we have drawn the specific value $x=x_0$, we would
like to test the hypothesis ${\cal H}_f$ that the probability
distribution associated with the random variable $x$ is $f(x)$.

In order to do so we define the distribution function
\footnote{The distribution function is defined in such a way the the
  probability that $x\in [a,b]$ is $F(b)-F(a)$.}
$F(x)\equiv\int_{-\infty}^x f(x^{\prime}){\rm d}x^{\prime}$.  For
Eq.~(\ref{f(x)}),
\begin{equation}
  \label{F(x)}
  F(x) = \left\{ 
    \begin{array}{ll} 
      0 & {\rm if}~x\le 0, \\
      x & {\rm if}~x\in [0,1], \\
      1 & {\rm if}~x\ge 1. 
    \end{array} \right. 
\end{equation}
We then compare $F(x)$ with the best possible guess for a distribution function 
$F_{\rm guess}(x)$ that can be obtained given that $x=x_0$ has been ``drawn,'' 
namely,
\begin{equation}
F_{\rm guess}(x)=\theta(x-x_0).
\end{equation}
Note that it is very easy to generalize this to $N$ random drawings of
$x$, which yield, say, $x_0,x_1,\ldots,x_{N-1}$ \cite{KS_ref}.

The (two-sided) KS statistic (``$D$-function'') is defined by \cite{KS_ref}
\begin{equation}
  \label{D}
  D={\rm sup}_x[|F_{\rm guess}(x)-F(x)|].
\end{equation}
In the example we have been discussing, $D_0=x_0$ if $x_0\ge 0.5$ or
$D_0=1-x_0$ if $x_0\le 0.5$ (note that the two expressions agree at
$x_0=0.5$, and we assume that $x_0\in[0,1]$).  
If the hypothesis ${\cal H}_f$ is correct, the
probability that a larger value of $D$ ({\it i.e.} a ``worse fit'')
would be computed from a different random drawing of $x$ is \cite{KS_ref} 
\begin{equation}
  \label{P_KS}
  P(D\ge D_0)=2(1-D_0),
\end{equation}
which is, in the example we have been discussing,
\begin{equation}
  P(x_0) = \left\{ 
    \begin{array}{ll} 
      2x_0 & {\rm if}~x_0\le 0.5, \\
      2(1-x_0) & {\rm if}~x_0\ge 0.5. 
    \end{array}
  \right.
\end{equation}
The smaller the value of $P(x_0)$, the less likely it is that ${\cal
  H}_f$ is correct. In this context, we allow statements such as
${\cal H}_f$ is only allowed at the $[1-P(x_0)]$ confidence level.

%{\sl Test with the Quark and the Leptonic Mixing Matrices}--
We wish to apply the test described above to the MNS and CKM mixing
matrices for leptons and quarks, respectively.  Our model is that the
mixing matrices are random variables drawn from a ``flat''
distribution of unitary $3\times 3$ matrices. Following the PDG
convention, we define the three mixing angles as in Table~\ref{table}.
\begin{table}
  \caption{$\sin^2\theta_{ij}$ in the MNS and CKM mixing matrices, according to
           the PDG parameterization \cite{PDG}. In square brackets we quote the
           currently allowed experimental values for the CKM (MNS) entries
           at the 90\% (three sigma) confidence level.}
  \label{table}
  \begin{tabular}{|c|c|c|} \hline
    ``angle'' & CKM [90\% expt.] & MNS [3$\sigma$ expt.] \\ \hline
    $\sin^2\theta_{13}$ & $|V_{ub}|^2~~[(6.2-23)\times10^{-6}]$ &
    $|U_{e3}|^2~~[0-0.05]$ \\ \hline 
    $\sin^2\theta_{12}$ & $\sin^2\theta_C~~[0.048-0.051]$ &
    $\sin^2\theta_{\rm sol}~~[0.2-0.5]$ 
    \\ \hline
    $\sin^2\theta_{23}$ & $|V_{cb}|^2~~[(1.4-1.9)\times10^{-3}]$ &
    $\sin^2\theta_{\rm atm}~~[0.35-0.65]$ 
    \\ \hline  
  \end{tabular}
\end{table}
Within this convention, the hypothesis is that the marginalized
probability density function is given by (see \cite{anarchy2} for a
detailed discussion of this point)
\begin{equation}
  \int f({U(3)}){\rm d(phases)}=
  f(\cos^4\theta_{13},\sin^2\theta_{12},\sin^2\theta_{12})=1,
\end{equation} 
where we have integrated over all (both physical and unphysical)
complex phases. The mixing angles are defined such that
$\theta_{ij}\in[0,\pi/2],~\forall i,j$.  The probability distribution
is flat in $\sin^2\theta_{12}$, $\sin^2 \theta_{23}$, and $\cos^4
\theta_{13}$.  It is clear that $f=1$ is correctly normalized,
\begin{equation} 
\int f~
{\rm d}(\cos^4\theta_{13}){\rm d}(\sin^2\theta_{12})
{\rm d}(\sin^2\theta_{23})=1, \label{eq:f}
\end{equation}
as it should be.

Since anarchy implies that the three mixing angles are distributed
as {\it uncorrelated}\/ random variables according to Eq.~(\ref{eq:f}), 
we are allowed to perform a {\it separate} KS test for
each of the three mixing angles. The three distinct $D$-functions are
(from Eq.~(\ref{D}) and the line that succeeds it),
\begin{eqnarray}
  D_{\theta^0_{13}}&=&(1-\sin^2\theta^0_{13})^2, \\
  D_{\theta^0_{12}}&=&1-\sin^2\theta^0_{12}, \\
  D_{\theta^0_{23}}&=&1-\sin^2\theta^0_{23}.
\end{eqnarray}
The superscript $0$ refers to the randomly picked value ({\it i.e.,}\/
the physical value, ``drawn'' by Nature) of the corresponding mixing
angle. We have assumed that $\sin^2\theta^0_{12,23}<1/2$,
$\cos^4\theta^0_{13}>1/2$. The generalization for all values of
$\theta_{ij}^0$ is trivial and does not add to our discussion
\footnote{Of course, all angles in the quark sector satisfy
  $\sin^2\theta_{ij}\ll 1/2$. The same is true of $|U_{e3}|^2$, while 
  the preferred values of
  $\sin^2\theta_{\rm sol}$ are smaller than 1/2 at the three sigma level. 
  We don't know whether $\sin^2\theta_{\rm atm}$ is
  less than or greater than 1/2. The experimental information we do
  have is such, however, that $\theta_{\rm atm}$ and
  $\pi/2-\theta_{\rm atm}$ cannot be discriminated. These ``degenerate''
  solutions lead to the same $D_{\theta^0_{23}}$.}.

Because the three ``random variables'' are not correlated, we
calculate the probability that a different random draw would yield a
worse result by computing the area where the product of three
one-variable probabilities
\begin{eqnarray}
  \epsilon^0&=&P(\theta^0_{12}) \times P(\theta^0_{13}) \times
  P(\theta^0_{23}) \nonumber \\
  &=& 8\sin^2\theta^0_{12}(2\sin^2\theta^0_{13}-\sin^4\theta^0_{13})
  \sin^2\theta^0_{23}
\end{eqnarray}
is worse than the data.  Here,
$P(\theta^0_{ij})=2(1-D_{\theta^0_{ij}})$, as in 
Eq.~(\ref{P_KS}). Therefore
\begin{equation}
  P(KS) = \int {\rm d}(\cos^4 \theta_{13}) {\rm d}(\sin^2 \theta_{12}) 
  {\rm d}(\sin^2 \theta_{23}) \theta(\epsilon-\epsilon^0),
\end{equation}
where $\epsilon$ is given by the same expression as $\epsilon^0$,
evaluated at the observed values of the mixing parameters, and 
$\theta(\epsilon-\epsilon^0)$ is the usual step function. We obtain
\begin{equation}
  P(KS) = \epsilon\left(1-\log\epsilon+\frac{1}{2}\log^2\epsilon\right).
\end{equation}

By using the best fit values $\sin^2\theta_{12} = 0.3$ \cite{Fogli:2002pb}
and $\sin^2\theta_{23} = 0.5$ \cite{atm} for the MNS matrix, we find
\begin{equation}
  \label{eq:2}
  \epsilon = 2.4 (\sin^2\theta_{13} - \frac{1}{2} \sin^4\theta_{13}).
\end{equation}
Given the bound $\sin^2\theta_{13} \lesssim 0.05$, the anarchical
hypothesis is consistent with the current data, with probability
$64\%$.

One can also check whether anarchy works in the quark sector. Using
the values tabulated in Table~\ref{table}, one obtains a probability
smaller than $6\times 10^{-6}$, implying that the hypothesis that the
CKM matrix is a random unitary $3\times3$ matrix is safely discarded. 
Hence, a
fundamental distinction among the three flavors of quarks seems to be required.

Once we have established as consistent the hypothesis that the MNS
matrix is a matrix drawn from a random sample of unitary $3\times 3$
matrices, we now turn the argument around, and try to place a lower
limit on $\theta_{13}$.  What we require is that $P(KS) > 1-P_0$,
where $P_0$ is defined to be the confidence level of the limit.

\begin{figure}[t]
%\vspace{1cm}
  \centerline{\includegraphics[width=\columnwidth]{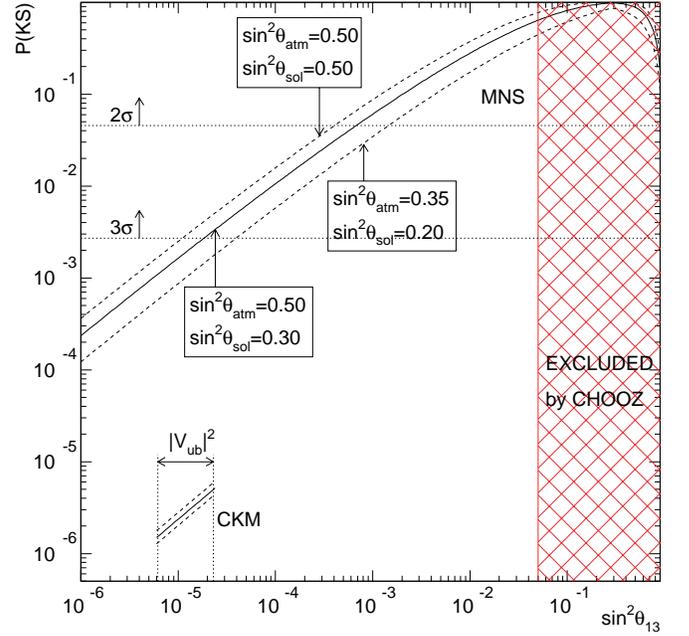}}
  \caption{$P(KS)$ for the MNS matrix
    as a function of $\sin^2\theta_{13}\equiv|U_{e3}|^2$, 
    see text for details.  The
    top dashed curve corresponds to
    $\sin^2\theta_{\rm sol}=\sin^2\theta_{\rm atm}=0.5$, 
    while the bottom dashed
    curve corresponds to $\sin^2\theta_{\rm sol}=0.2$ and
    $\sin^2\theta_{\rm atm}=0.35$. The solid curve corresponds to the best
    fit values $\sin^2\theta_{\rm sol}=0.3$ and $\sin^2\theta_{\rm atm}=0.5$.
    The hatched region is currently excluded by the neutrino data.
    In the bottom left corner, $P(KS)$ for the CKM matrix as a function
    of $\sin^2\theta_{13}\equiv|V_{ub}|^2$ is also depicted within the
    experimentally allowed range for $|V_{ub}|^2$, assuming
    that the values of $|V_{cb}|^2$ and $\sin^2\theta_C$ vary within
    the range indicated in Table~\ref{table}.}
  \label{KS_prob}
\end{figure}

Fig.~\ref{KS_prob} depicts $P(KS)$ for the MNS matrix 
as a function of $\sin^2\theta_{13}\equiv |U_{e3}|^2$ 
within the three sigma bounds
allowed experimentally for $\sin^2\theta_{12}
\equiv\sin^2\theta_{\rm sol}$ and $\sin^2\theta_{23}
\equiv\sin^2\theta_{\rm atm}$, as tabulated in Table~\ref{table}. For
the best fit values of $\sin^2\theta_{\rm atm}$ and $\sin^2\theta_{\rm sol}$,
one is able to ``rule out'' $|U_{e3}|^2<0.0007$ at the two sigma level
and $|U_{e3}|^2<0.00002$ (which is, curiously, the 
upper bound for $|V_{ub}|^2$) at the three sigma level.  
Fig.~\ref{KS_prob} also depicts $P(KS)$ as a function of
$\sin^2\theta_{13}\equiv |V_{ub}|^2$ for the CKM matrix within the
90\% experimentally allowed ranges defined in Table~\ref{table}.

Note, however, that these bounds are obtained {\it a posteriori}\/, and
turn out to be rather weak. This is due to the fact that the observed
values of the angles $\theta_{12}$ and $\theta_{23}$ agree ``too
well'' with the anarchical hypothesis, hence allowing a
larger-than-usual fluctuation for $\theta_{13}$.  We believe that
$P(KS)$ is a good tool for testing the anarchical hypothesis against
the data, but not as useful a tool for studying what values of
$|U_{e3}|^2$ are preferred by the hypothesis.  For this reason, we
choose to make use of another method of obtaining lower bounds on
$|U_{e3}|^2$ assuming anarchy.  This method can be thought of as
yielding an {\it a priori}\/ prediction for $|U_{e3}|^2$, which, we
believe, is more appropriate, and will be described promptly.

%% There are other ways of obtaining a lower bound on $|U_{e3}|^2$ given
%% the anarchical hypothesis, and we would like to mention two of them.

Let us compute the marginalized probability density
function of $\theta_{13}$ {\sl only.}\/ From Eq.~(\ref{eq:f}), 
\begin{equation}
\int f~{\rm d}(\sin^2\theta_{12})
{\rm d}(\sin^2\theta_{23})\equiv g(\cos^4\theta_{13})=1.
\label{prob_1d}
\end{equation}
This probability distribution ($g$) is rather similar to $f$, (see
Eq.~(\ref{eq:f})) but is to be interpreted in a slightly different
way. Remember that $f$ is the probability distribution function
derived for $U(3)$ marginalized over all CP-odd phases, including:
three unphysical phases that can be removed by redefining fermionic SM
fields, two ``Majorana phases'' which may or may not be physical,
depending on whether or not the neutrinos are Majorana fermions, and
one ``Dirac phase.'' Marginalyzing over unphysical phases is the only
reasonable procedure to follow, and we have chosen to also marginalize
over the physical phase(s) in order to address whether the current
information we have on mixing angles fits the anarchical hypothesis.
This {\sl choice}\/ only makes sense if the anarchical hypothesis for
the ``whole'' MNS matrix is consistent, which is the case as we have
observed above.  Similarly, $g$ is the probability distribution
function derived for $U(3)$ marginalized over all CP-odd phases {\sl
  and}\/ the ``solar'' and ``atmospheric'' mixing angles.
%% One can choose to do this in order to check whether the current information
%% we have on $|U_{e3}|^2$ fits the anarchical hypothesis. it is important to 
%% emphasize that making such a choise is only consistent 
%% if the anarchical hypothesis for the ``entire'' unitary mixing matrix 
%% is consistent.   

The single variable probability is
\begin{equation}
P(KS)_{1d}=P(\theta_{13}^0)
     =4(\sin^2\theta_{13} - \frac{1}{2} \sin^4\theta_{13}). 
\label{P(KS)_1d}
\end{equation}

\begin{figure}[t]
%\vspace{1cm}
  \centerline{\includegraphics[width=\columnwidth]{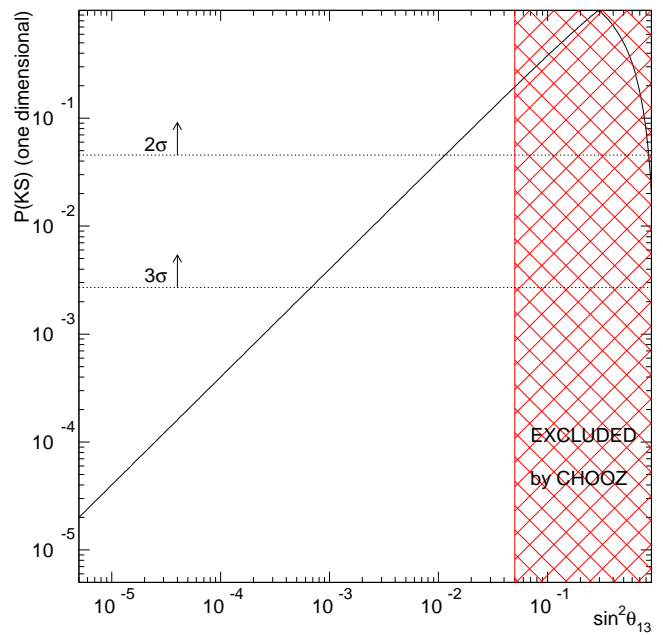}}
  \caption{$P(KS)_{1d}$ as a function of $\sin^2\theta_{13}\equiv|U_{e3}|^2$, 
    see text for details.  
    The hatched region is currently excluded by the neutrino data.}
    \label{KS_prob_1d}
\end{figure}

Fig.~\ref{KS_prob_1d} depicts $P(KS)_{1d}$ as a function of 
$\sin^2\theta_{13}\equiv|U_{e3}|^2$. Again, we 
can interpret $P(KS)_{1d}$ as the probability that a 
``worse fit'' is obtained assuming
that $\theta_{13}$ is a random variable drawn from the 
probability distribution Eq.~(\ref{prob_1d}). 
These bounds are {\it a priori}\/ predictions of the anarchy unlike
the previous ones, and should be taken more seriously.
%% As before, we allow ourselves to 
%% state that this hypothesis is, given the current bound $|U_{e3}|^2<0.05$,
%% consistent, with probability 20\%. 
%% Similar to the three-variable case, we
We exclude $|U_{e3}|^2<0.011$ (0.0007) at the two (three) sigma
confidence level. Note that the CKM-equivalent $P(KS)_{1d}=P(V_{ub})$ is
less than $10^{-4}$ (this can be easily read off from Fig.~\ref{KS_prob_1d}),
again indicating that the anarchical hypothesis in the quark sector can
be safely discarded.

%%%%
%One can also make use of $g$ (Eq.~(\ref{prob_1d})) 
%directly in order to set a lower bound 
%on $|U_{e3}|^2$. Given that the anarchical hypothesis is consistent,
%we know the probability distribution for $|U_{e3}|^2$, namely,
%$g$. One
%can straightforwardly compute the probability that 
%$|U_{e3}|^2>|U_{e3}|^2_{\rm min}$ as 
%\begin{eqnarray}
%&P(|U_{e3}|^2>|U_{e3}|^2_{\rm min})\equiv \frac{ 
%\int_{(0.95)^2}^{(\cos^4\theta_{13})_{\rm max}}{\rm d}(\cos^4\theta_{13})}
%{N}, \label{direct_bound} \\
%&N=\int_{(0.95)^2}^{1}{\rm d}(\cos^4\theta_{13})=0.0975,
%\end{eqnarray}
%where $(\cos^4\theta_{13})_{\rm max}=(1-|U_{e3}|^2_{\rm min})$, and
%$(0.95)^2$ is the lower limit on $\cos^4\theta_{13}$ imposed by CHOOZ.
%Note that we have rescaled the probability distribution by forcing it
%to the ``physical region,'' dictated by the current upper bound on
%$|U_{e3}|^2$. From Eq.~(\ref{direct_bound}), we constrain
%$|U_{e3}|^2>0.0022$ (0.00015) at the two (three) sigma confidence
%level.  {\bf This bound is a mixture of {\it a posteriori}\/ and {\it a
%  priori}\/ arguments; it is {\it a posteriori}\/ because it is
%one-sided motivated by the fact that the data prefers $|U_{e3}|^2$ to
%be small, while it is {\it a priori}\/ because it does not depend on
%the observed values of $\theta_{12}$ and $\theta_{23}$.}
%
%{\bf Andr\'e, I actually don't like this bound.  It basically assumes
%that the ``physical region'' of $\cos^4 \theta_{13} > 0.05$ as given,
%and defines the probability only inside the region.  It not only makes
%the bound artificially weak, but also I find it somehow ``wrong.''}
%%%

Finally, It is worth recalling that anarchy predicts a flat probability
distribution for the CP-violating phase $\delta$ \cite{anarchy2}, and
hence the distribution in $\sin\delta$ is $1/|\cos\delta|$, peaked at
$\sin\delta=\pm 1$.  If anarchy is correct, chances are that the
observation of CP violation in long-baseline oscillation experiments
is indeed within reach!

We now summarize our results, with more discussions to follow. We have
statistically tested the hypothesis that the MNS matrix is a matrix
drawn from a random ``flat'' sample of unitary $3\times 3$ matrices.
According to the KS test performed, this ``anarchical hypothesis'' is
consistent with the data.  The anarchical hypothesis fails the KS test
when it is performed with the CKM matrix. Our result is different from
other attempts to statistically ``test'' anarchy. For example, the
authors of \cite{no_anarchy} have claimed that the neutrino sector
prefers the existence of some symmetry behind neutrino masses and
mixing angles to completely random entries. We have not attempted to
perform such a ``compartive test,'' which is, at least, hard to
interpret in a well defined way. We do not believe that such tests are
capable of indicating whether one hypothesis is favored with respect
to the other. Our test has a well defined statistical interpretation,
and directly probes whether anarchy in the neutrino sector is a good
hypothesis.

Having checked that anarchy is consistent with our current
understanding of the MNS matrix, we were able to use the anarchical
hypothesis to ``predict'' the value of the still unobserved mixing
angle $\theta_{13}$. At the two sigma level, anarchy requires that
$\sin^2\theta_{13}>0.011$, for example (bound obtained from $P(KS)_{1d}$,
see Fig.~\ref{KS_prob_1d}). 
If there is indeed no
structure in the leptonic mixing matrix, it seems very likely that one
should be able to observe CP-violation in long-baseline neutrino
oscillation experiments, as not only are all angles large, but the
CP-odd parameter $\sin\delta$ is also ``predicted'' to be large.

We have nothing to say about the value of the neutrino masses.  The
hypothesis we tested is that the MNS matrix is ``random,'' independent
of whether the masses are degenerate, partially degenerate or
hierarchical \cite{anarchy2}.  Even in the case of non-LMA solutions
to the solar neutrino puzzle (currently ruled out at
99.95\% C.L. \cite{KamLAND}), one can obtain random mixing
matrices \cite{min_model}. 
Incidently, it is interesting to note that the the neutrino
masses seem to be ``less hierarchical'' than the charged fermion
masses.  Assuming that the neutrino masses are not degenerate, it
turns out that $m_3/m_2\simeq \sqrt{\Delta m^2_{23}/\Delta m^2_{12}}=
3-13$, not too far away from unity (of course, we do not know
$m_2/m_1\ldots$).  This is consistent with random mass matrices
generated via the seesaw mechanism \cite{anarchy1}.

We would like to underline important assumptions and limitations of
our result. By hypothesis, the probability distributions for the
mixing angles are uncorrelated.  Our discriminatory procedure does
not include information regarding whether the different variables are
more likely to be correlated than not. Given the minimal statistics
(provided by the fact that we live in only one Universe), adding this
sort of information would not lead to different conclusions, although
one should start to worry if, say, it turns out that
$\sin^22\theta_{23}=\sin^22\theta_{12}=1$.  One should also be warned
that the KS test performed here need not be the most powerful test for
the anarchical hypothesis, statistically speaking \cite{KS_ref}.
    
Finally, we emphasize what our result {\it does not}\/ imply.
Although the anarchical hypothesis is consistent with the 
data, neutrino mass
models which rely on flavor symmetries and nontrivial ``textures'' are
not disfavored in any well defined way. Some are perfectly justified
by top-down arguments, including, say, grand unification of matter
fields. We would like to point out, however, that the ``burden of
proof'' is with the models that assume that there is structure in the
leptonic mixing matrix. The anarchical hypothesis may be viewed as the
simplest of flavor models -- a model of flavor without flavor. In
light of our long experience with quark masses and mixing angles, it
is remarkable that, in the neutrino sector, one can do without new
symmetry principles in order to appreciate the entries of the MNS
mixing matrix.

{\bf Note Added --}\/ After the first version of the this manuscript
became publicly available, a preprint discussing our results
\cite{Espinosa} appeared. All of the comments contained there apply to
this version of our manuscript as well.  While we appreciate most of
the arguments contained in \cite{Espinosa}, we disagree with its
author in a few key points. Most importantly, we do not agree with the
claim that $\sin^2\theta_{12},\sin^2\theta_{23},\cos^4\theta_{13}\in
[0,1]$ are ``angular variables.'' Note that the Haar measure is flat
in these variables rather than in the mixing angles, which are
convention dependent.  Therefore we stand by our claim that a KS test
can be used to test the anarchical hypothesis. Furthermore,
\cite{Espinosa} contains an alternative statistical test of the
anarchical hypothesis (a Kuiper's test, see \cite{Espinosa} for
details), and the result that maximal mixing is ``preferred'' is
obtained, in qualitative agreement with the results presented here.
The author of \cite{Espinosa}, however, dismisses the result of the
Kuiper's test, claiming that the statistical sample is too small.
While we appreciate that some are uneasy about the statistics of very
small data samples, namely one chosen by Mother Nature, we point out
that the dismissal of such results is not mathematically justified.

\acknowledgments We thank Paolo Creminelli for very useful questions
and discussions, and Jos\'e Ramon Espinosa for corresponce regarding
\cite{Espinosa}.  The work of AdG is supported by the US Department of
Energy Contract DE-AC02-76CHO3000.  The work of HM is supported in
part by the Director, Office of Science, Office of High Energy and
Nuclear Physics, Division of High Energy Physics of the U.S.
Department of Energy under Contract DE-AC03-76SF00098 and in part by
the National Science Foundation under grant PHY-00-98840.

\end{document}